\documentclass[aps,prb,superscriptaddress,reprint,showpacs,twocolumn,
preprintnumbers,amsmath,amssymb,floatfix]{revtex4-1}
\usepackage{graphicx}
\usepackage{color}
\usepackage[latin1]{inputenc}
\usepackage{amssymb}
\usepackage{bm}
\usepackage{stmaryrd}
\usepackage{wasysym}
\usepackage{subfigure}
\usepackage{afterpage}

\newcommand{\asnmr}{$^{75}$As }
\newcommand{\tsl}{$1/T_1$ }
\newcommand{\tss}{$1/T_2$ }

\newcommand{\bafeas}{Ba(Fe$_{1-x}$Rh$_x$)$_2$As$_2$ }

\begin{document}

\title{Effect of proton irradiation on the normal state low-energy excitations of
Ba(Fe$_{1-x}$Rh$_x$)$_2$As$_2$ superconductors}

\author{M. Moroni}
\email{matteo.moroni01@universitadipavia.it}
\affiliation{Department of Physics, University of Pavia-CNISM,
I-27100 Pavia, Italy}
\author{L. Gozzelino}
\affiliation{Istituto Nazionale di Fisica Nucleare, Sez. Torino, Torino 10125,
Italy}
\affiliation{Politecnico di Torino, Dept. of Applied Science and Technology, 10129 Torino, Italy}
\author{G. Ghigo}
\affiliation{Istituto Nazionale di Fisica Nucleare, Sez. Torino, Torino 10125,
Italy}
\affiliation{Politecnico di Torino, Dept. of Applied Science and Technology, 10129 Torino, Italy}
\author{M.~A.~Tanatar}
\affiliation{Ames Laboratory USDOE and Department of Physics and
Astronomy, Iowa State University, Ames, Iowa 50011, USA}
\author{R. Prozorov}
\affiliation{Ames Laboratory USDOE and Department of Physics and
Astronomy, Iowa State University, Ames, Iowa 50011, USA}

\author{P.~C. Canfield}
\affiliation{Ames Laboratory USDOE and Department of Physics and
Astronomy, Iowa State University, Ames, Iowa 50011, USA}

\author{P. Carretta}
\affiliation{Department of Physics, University of Pavia-CNISM,
I-27100 Pavia, Italy}

\begin{abstract}
We present a \asnmr Nuclear Magnetic Resonance (NMR) and resistivity study of the
effect of 5.5 MeV proton irradiation on the optimal electron doped
($x=$ 0.068) and overdoped ($x=$ 0.107)
Ba(Fe$_{1-x}$Rh$_x$)$_2$As$_2$ iron based superconductors. While the proton induced defects only mildly suppress the critical temperature and increase residual resistivity in both compositions, sizable broadening of the NMR spectra was observed in all the irradiated samples at low
temperature. The effect is significantly stronger in the optimally doped
sample where the Curie Weiss temperature dependence of the line
width suggests the onset of ferromagnetic correlations coexisting
with superconductivity at the nanoscale. 1/T$_2$ measurements
revealed that the energy barrier characterizing the low energy
spin fluctuations of these compounds is enhanced upon proton
irradiation, suggesting that the defects are likely slowing down the  fluctuations between
($0,\pi)$ and ($\pi$,0) nematic ground states.
 \end{abstract}


\maketitle


\section{\label{sec:intro} Introduction}
Chemical substitution is the most common approach used to
introduce impurities in strongly correlated electron systems in
order to probe their local response function. However, this method
often gives rise to structural distortions, unwanted
inhomogeneity and to charge doping. Accordingly, in order to
study the effect of the bare impurities the right dopant must be
carefully chosen and the options are often very limited. Thus irradiation with 
energetic particles, electrons and ions, may represent a powerful alternative 
to chemical
substitutions.  Radiation induced defects have been extensively
employed in high temperature superconductors to investigate the
pair breaking effect of non magnetic scattering centers and to
study the pinning of the Abrikosov vortices. In particular, heavy
ions irradiation (e.g. with Au and Pb) induces strongly
anisotropic columnar defects, which are effective in
pinning the flux vortices~\cite{massee2015,tamegai2012}.
Conversely, low mass ions, such as protons, $\alpha$ particles or
electrons, give rise to uniformly distributed point like defects
whose density can be precisely controlled. In the cuprates the
decrease of the superconducting transition temperature T$_c$ with
the radiation fluence $\phi$ was found to strongly depend on the
ion type, on its energy and on the total dose~\cite{sommers89}.
Remarkably, in YBa$_{2}$Cu$_{3}$O$_{7-\delta}$ and
Tl$_{2}$Ba$_{2}$CuO$_{6+x}$, it was found~\cite{rullier2000} that
the defects introduced by electron irradiation play a role
analogous to nonmagnetic Zn impurities and the magnitude of
$d$T$_c$/$d\phi$ is consistent with the theoretical prediction for
a $d$-wave superconductor~\cite{radke93}.

In the iron based superconductors (IBS) several irradiation
studies have been conducted with
heavy~\cite{prozorov2010,salovich2013,ghigo2014,civale2014,
ohtakea2013,kilstrom2014,tamegai2012,ghigo2013}ions, light
ions~\cite{nakajima2010,taen2013,civale2014,kilstrom2014} and electrons \cite{MatsudaBaP,ProzorovPRX,VanderBeek, KChoBaKirradiation,KChoScienceAdvance,FeSeTeknowijoyo}. In
these compounds T$_c$ suppression by radiation damage is rather weak for 
optimally doped compositions but becomes stronger in under-doped and overdoped 
compositions. 
Simultaneous studies of $T_c$ suppression and London penetration depth as a function of doping in Ba$_{1-x}$K$_x$Fe$_2$As$_2$  \cite{ProzorovPRX,KChoScienceAdvance} conclude that both quantities can be reasonably fit to $s^\pm$
model~\cite{mazin2008,chubukov2008} which is the leading
candidate for describing the pairing state in most of the IBS 
\cite{fernandes2012,bang2017,ghigo2017}.
Interestingly, these results are consistent with the reduced T$_c$
suppression induced by non-magnetic Zn doping in
Ba(Fe$_{1-x}$Co$_{x}$)$_{2}$As$_{2}$ and LaFeAsO$_{1-x}$F$_x$
\cite{li2012,li2009,li2010,guo2010}. This weak effect of
diamagnetic impurities in IBS is not necessarily an indication of
a different gap symmetry. In fact one should notice that the defects
weaken also the spin density wave (SDW) phase competing with
superconductivity (SC) in the underdoped part of the phase
diagram\cite{fernandes2012,KimBaK,ReidBaK}. Hence, $d$T$_c$/$d\phi$
strongly depends on the system parameters in the underdoped
regime, both for proton irradiation
and nonmagnetic Zn
doping~\cite{li2012,li2009,li2010,guo2010,nakajima2010,taen2013}.

The studies cited above focus mainly on the superconducting state
and no reports can be found in the literature on a systematic
investigation of the effects of irradiation on the normal state
properties of IBS, in particular on the spin and nematic
correlations~\cite{chubukov2016}. In 122 iron based
superconductors very slow spin fluctuations have been detected
above T$_c$ with Nuclear Magnetic Resonance (NMR)
~\cite{curro2009,xiao2012,curro2015} and have been ascribed to nematic
fluctuations among (0,$\pi$) and ($\pi$,0) correlated regions \cite{kissikov2016}.
Recently $^{75}$As \tss NMR measurements in the electron doped
\bafeas revealed~\cite{lucia2013,lucia2016} that these fluctuation
are not only present in the underdoped part of the phase diagram
but extend up to at least 11\% Rh doping, well into the overdoped
regime.

In this manuscript we show that proton induced defects
significantly affect the slow spin fluctuations revealed by $^{75}$As
\tss in Ba(Fe$_{1-x}$Rh$_x$)$_2$As$_2$, suggesting that the fluctuations developing between $(0,\pi$) and ($\pi$,0) phases are affected by the
disorder. Moreover, we observe a broadening of the $^{75}$As NMR
spectra induced by proton irradiation and  for the optimally doped
0.068 Rh sample it evidenced that the defects induce
ferromagnetically correlated regions around the impurities,
coexisting with superconductivity.

\section{\label{sec:methods} Experimental methods and results}

\begin{figure}[t!]
\includegraphics[width=8.5cm, keepaspectratio]{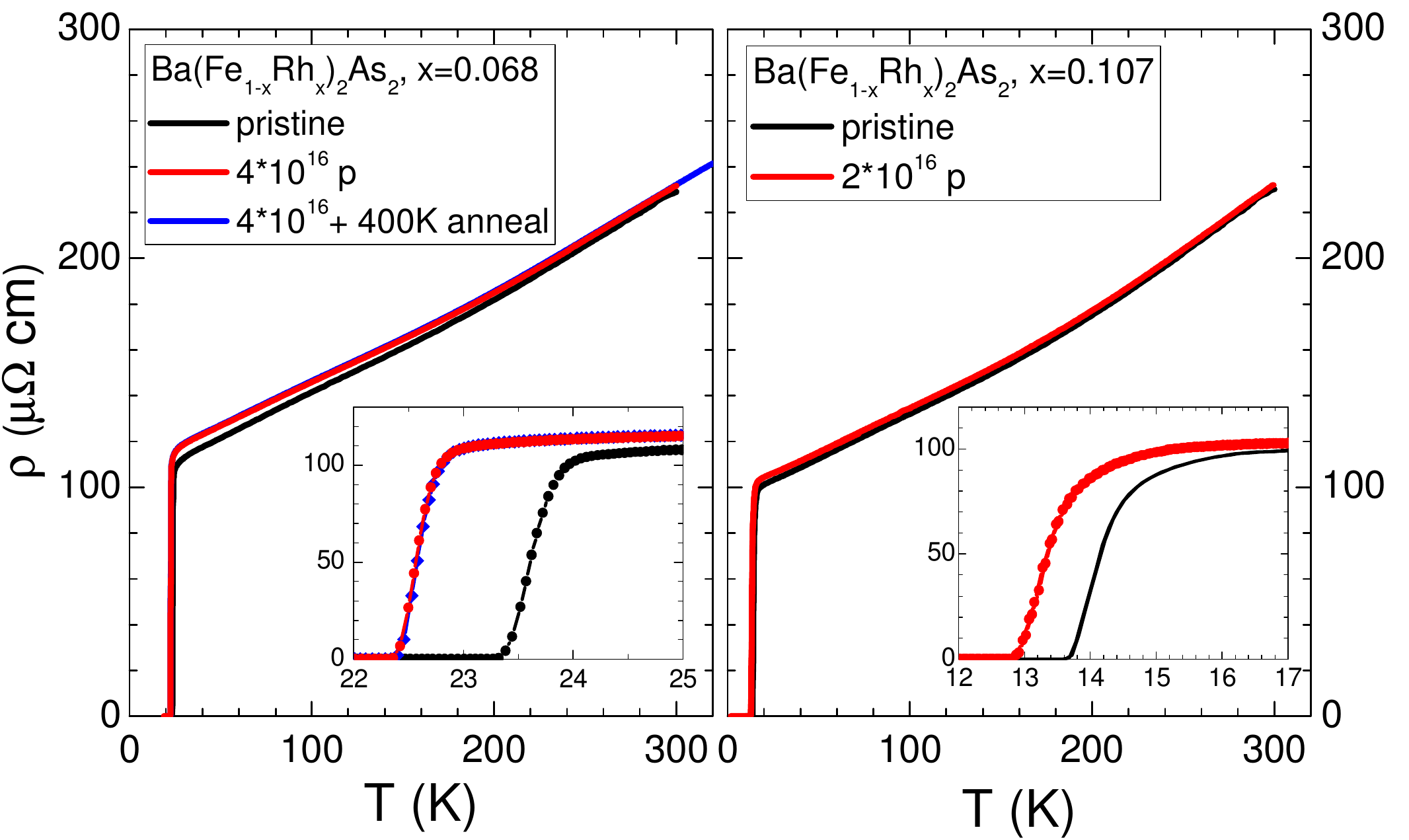} 
\caption{(Color online) Temperature-dependent electrical resistivity $\rho (T)$ of optimally doped $x$=0.068 (left panel) and overdoped $x$=0.107 (right panel) samples of Ba(Fe$_{1-x}$Rh$_x$)$_2$As$_2$. Insets zoom into superconducting transition range. Black lines show $\rho(T)$ for samples in the pristine state, red lines show the data for the same samples after proton irradiation. Blue line in left panel shows $\rho (T)$ of the same sample after annealing at 400~K, revealing permanent character of proton irradiation damage, in contrast to damage by electron irradiation \cite{ProzorovPRX}. Note non-parallel shift of the $\rho (T)$ curves after irradiation, revealing Matthiessen rule violation. } 
\label{rho}
\end{figure}

The measurements presented in this work were performed on \bafeas
single crystals with Rh content of $x=$ 0.068 (optimally doped
sample) and $x=$ 0.107 (overdoped sample). The crystals were grown
using the method described in
Ref.~\onlinecite{canfield-crystals}. The samples were then
characterized by means of resistivity and SQUID magnetometry
measurements. 
Electrical resistivity measurements were made using four-probe technique on 
cleaved samples with typical dimensions 2$\times$0.5$\times$0.05 mm$^3$, with 
long dimension corresponding to [100] crystallographic direction. Low 
resistance contacts to the samples were made by soldering 50 $\mu$m Ag wires 
using Sn \cite{anisotropy,SUST,patent}. Measurements were made on 6 samples of 
$x$=0.068 and 7 samples of $x$=0.107. In both cases resistivity of the samples 
at room temperature $\rho(300K)$ was 230$\pm30$ $\mu \Omega$cm, consistent 
within error bars with the results for Co-doped compositions of similar $x$ 
\cite{pseudogap}. Selected crystals of each batch were then
irradiated with 5.5 MeV protons at the CN Van de Graaff
accelerator of INFN-LNL (Istituto Nazionale di Fisica Nucleare -
Laboratori Nazionali di Legnaro, Italy).
Contacts to the samples remained intact during irradiation, thus eliminating 
uncertainty of geometric factor determination and enabling quantitative 
comparison of resistivity measurements. To minimize the heating of the
crystals under irradiation the proton flux was always limited to
10$^{12}$ cm$^{-2}$ s$^{-1}$. The irradiation with 5.5 MeV protons
produces random point defects and some defect nanoclusters, due to
elastic scattering of protons against the target nuclei. The
thickness of the crystals was much smaller than the proton
implantation depth, as calculated by the SRIM-2013
code~\cite{SRIM} using the Kinchin-Pease approach. This ensured a
homogeneous defect distribution in the superconductor, as
evidenced by Fig.~\ref{de/dx}c where the energy lost by protons due to elastic 
scattering is plotted as a function of the implantation depth. 
In \mbox{Table}~\ref{dpa} the 
average displacement damage and the inferred average distance between 
proton-induced point defects are reported as a function of the irradiation 
fluence. It has to be noted that this distance should be assumed as a lower 
limit since the primary point defects (Frenkel pairs) could migrate to form 
small clusters and some defects could anneal out. After crossing the whole 
crystals thickness protons get implanted into the sample-holder. 

\begin{figure}[t!]
	\includegraphics[width=8.5cm, keepaspectratio]{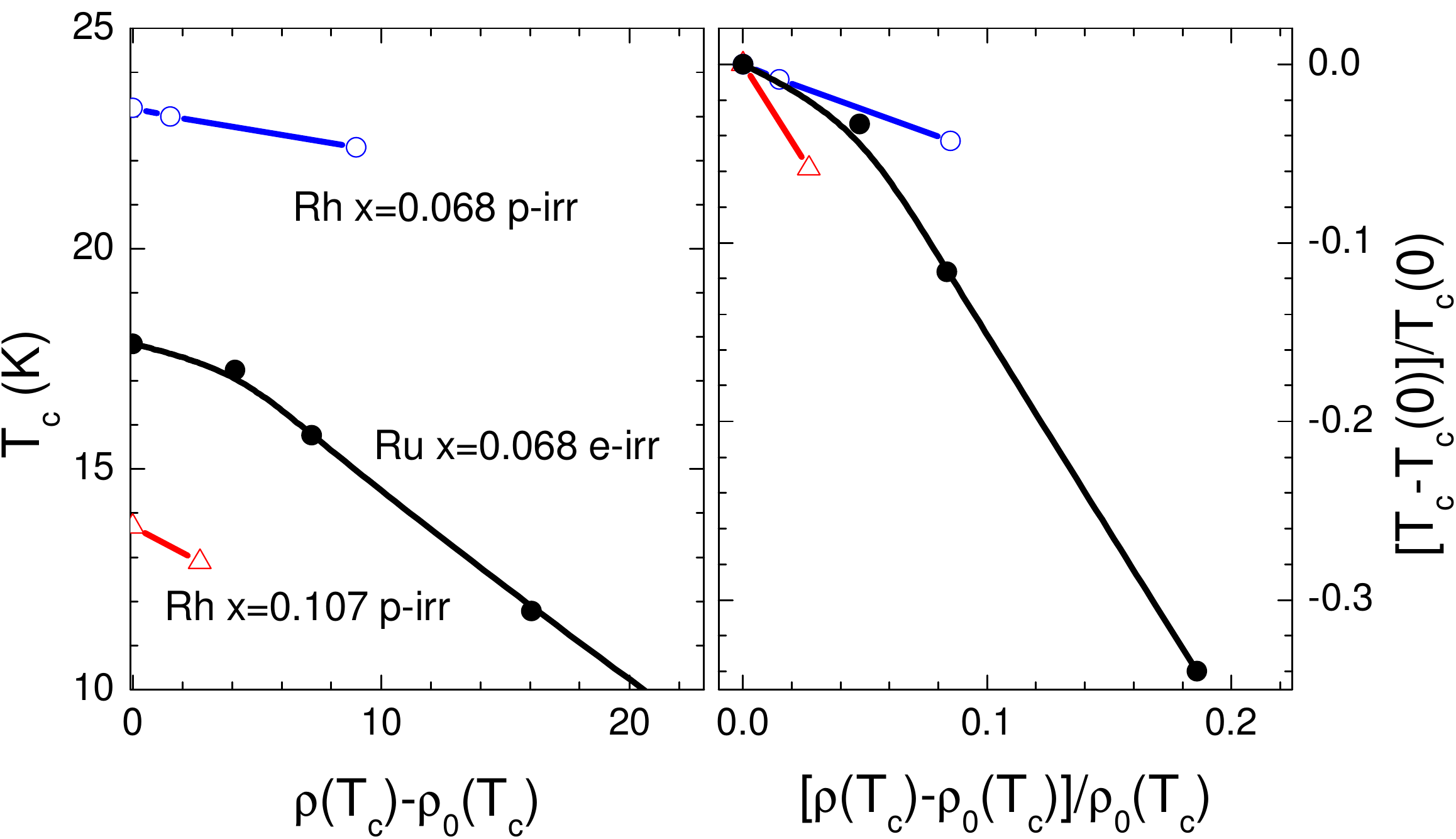} 
	\caption{(Color online) (Left panel) The superconducting transition temperature $T_c$ as a function of change in sample resistivity $\rho(T_c)$ for samples of Ba(Fe$_{1-x}$Rh$_x$)$_2$As$_2$ with optimal doping $x$=0.068 (blue curve, open circles) and $x$=0.107 (red curve, open up-triangles). For reference we show data for iso-electron substituted Ba(Fe$_{1-x}$Ru$_x$)$_2$As$_2$ at optimal doping $x$=0.24, subjected to low-temperature 2.5 MeV electron irradiation, Ref.~\onlinecite{ProzorovPRX}. Right panel shows same data plotted as a change in $T_c$ and resistivity $\rho(T_c)$ normalized by their values in pristine samples $T_c(0)$ and $\rho_0(T_c)$. } 
	\label{tc_rho}
\end{figure}
\begin{figure}[t]
	\includegraphics[width=8.5cm,
	keepaspectratio]{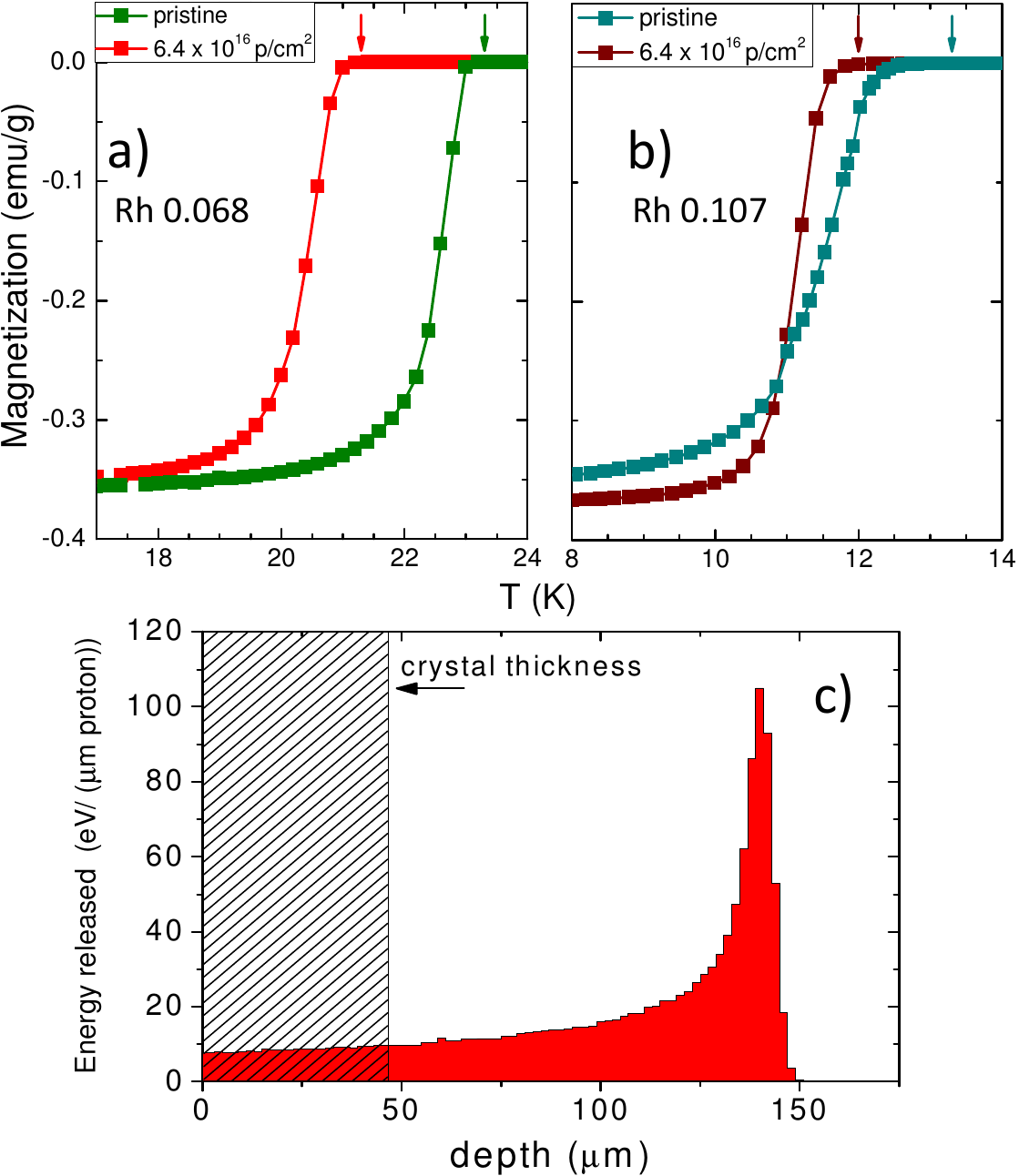} \caption{(Color online) a) and b): SQUID
		magnetization measurements for the $x=$0.068 sample (a) and $x=$
		0.107 sample (b) carried out before and after irradiation. The
		arrows indicate T$_c$ as determined by the onset of diamagnetism. 
		c): Distribution of the proton energy loss
		in the superconducting crystals (less than 50 $\mu$m thick) as a
		function of depth. The thickness of the thickest irradiated sample
		is about 45 $\mu$m, as evidenced in the picture. Therefore, the
		energy release can be considered homogeneous throughout the
		crystals, as well as the distribution of defects.} \label{de/dx}
\end{figure}
\begin{table}[b]
	\caption{Summary of the average displacements per atom (dpa) and distance between defects as a function of the proton irradiation fluences.
	}
	\label{dpa}
	\begin{ruledtabular}
		\begin{tabular}{c c c }
			$\phi$ (cm$^{-2}$) & dpa & Inter-defect distance (nm) \\
			\hline
			2$\times 10^{16}$ & 5.1$\times 10^{-4}$ & $3.5$  \\
			3.2$\times 10^{16}$ & 8.2$\times 10^{-4}$ & $3$  \\
			4$\times 10^{16}$ & 1$\times 10^{-3}$ & $2.8$  \\
			6.4$\times 10^{16}$ & 1.6$\times 10^{-3}$ & $2.4$  \\
		\end{tabular}
	\end{ruledtabular}
\end{table}
After irradiation the samples were again characterized with resistivity 
measurements to check the reduction of T$_c$ and $^{75}$As NMR measurements 
were then
carried out. Figure~\ref{rho} shows temperature dependent resistivity of the 
samples $x$=0.068 (left panel) and $x$=0.107 (right panel) before and after 
irradiation. Sample $x$=0.068 was subject to a fluence up to 4$\times$10$^{16}$ 
cm$^{-2}$, which resulted in approximately 1~K decrease of $T_c$ from 
23.3~K to 22.3~K as
determined by zero resistance criterion. Resistivity above the transition
increased from 106 to 115 $\mu \Omega$cm. To check the stability of irradiation
damage, sample of $x$=0.068 was heated up to 400~K. This protocol is known to 
show significant $T_c$ restoration and residual resistivity decrease in 
electron irradiated samples \cite{ProzorovPRX}, none of which is observed for 
proton irradiation. Due to a two times smaller irradiation fluence, 
2$\times$10$^{16}$ cm$^{-2}$, $T_c$ suppression in sample of $x$=0.107 is 
somewhat smaller, $\Delta T_c \approx$0.8~K, from 13.7 to 12.9~K. Resistivity 
increase is also notably smaller, $\Delta \rho \approx$3 $\mu \Omega$cm. 
It should be noticed that, for both compositions, the resistivity increase after irradiation is not a rigid offset as one would expect from Matthiessen rule. The shift becomes notably 
larger at low temperatures, in line with observations on hole-doped 
Ba$_{1-x}$K$_x$Fe$_2$As$_2$ \cite{KChoBaKirradiation}. 
In Fig.~\ref{tc_rho} we plot the effect of irradiation on $T_c$ as a function 
of the
residual resistivity change with respect to pristine sample 
$\rho(T_c)-\rho_0(T_c)$. In the right panel we plot the same data normalized by 
the values in pristine sample. For reference we plot the data for iso-electron 
substituted Ba(Fe$_{1-x}$Ru$_x$)$_2$As$_2$ at optimal doping $x$= 0.24, 
irradiated with 2.5 $MeV$ electrons \cite{ProzorovPRX}. The rates of $T_c$ 
variation are comparable in both cases, with some differences which can be 
ascribed to the variation of response due to the variation 
of doping level, rather 
than to the type of disorder. This is quite remarkable considering 
the very different 
annealing effect in the two cases. 

$T_c$ was also measured \textit{in situ} during the
NMR experiment by monitoring the detuning temperature of the NMR
probe resonating circuit. The decrease of T$_c$ after irradiation
($\phi=$3.2$\times 10^{16}$ cm$^{-2}$)
was found to be small both for the $x=$~0.068 (from 23.3 K before
irradiation to $\sim$22 K afterwards) and for the $x=$ 0.107 (from $\sim$13.3
K to $\sim$12.5 K). The samples were then irradiated again to increase the
total fluence to $\phi=$6.4$\times 10^{16}$ cm$^{-2}$, and SQUID (see 
Fig.~\ref{de/dx}a and \ref{de/dx}b) and NMR measurements were repeated. The 
expected displacement damage after these second irradiations and the 
corresponding average distance between proton-induced point defects are 
reported in \mbox{Table}~\ref{dpa}. The second irradiation
lowered T$_c$ to 21.3 K for $x=$ 0.068 and to 12 K for $x=$ 0.107.
Hence, the T$_c$ decrease rate is $d$T$_c$/$d\phi \simeq
$~0.3~$\times$10$^{-16}$~K$\cdot$cm$^2$ for the optimally doped
sample and about 0.2~$\times$10$^{-16}$~K$\cdot$cm$^2$ for the
overdoped one.

The values of $d$T$_c$/$d\phi$ are lower than those observed
in Ba(Fe$_{1-x}$Co$_{x}$)$_{2}$As$_{2}$ and
Ba$_{1-x}$K$_{x}$Fe$_{2}$As$_{2}$ irradiated with 3 MeV protons~\cite{nakajima2010,taen2013}.
This effect was expected since the non-ionizing energy loss, which
drives the number of defect produced per incoming proton,
decreases with increasing proton energy~\cite{sommers89}. This
means that, somewhat counterintuitively, the effectiveness of
protons in damaging the lattice decreases by increasing their
energy.

For each sample doping and dose value we measured the temperature
dependence of the $^{75}$As NMR linewidth, of the spin-lattice
relaxation rate (1/T$_1$) and of the spin-spin relaxation rate
(1/T$_2$). The magnetic field \mbox{\textbf{H}$_0$~$=7$~T} was
applied along the crystallographic $c$ axis unless otherwise
specified.

\begin{figure}[t]
\includegraphics[width=8.6cm,
keepaspectratio]{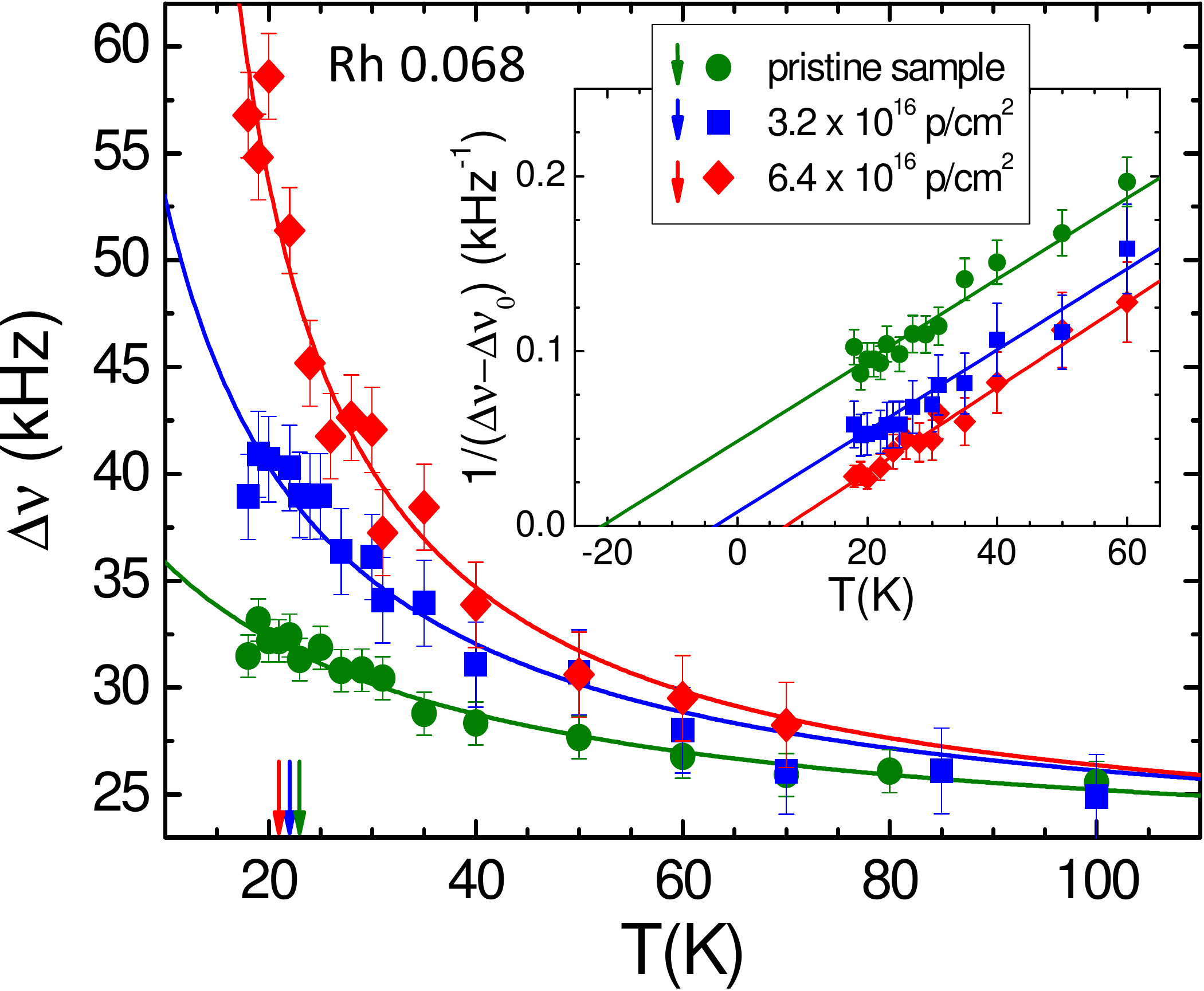} \caption{(Color online) Temperature
dependence of the Full Width at Half Maximum $\Delta\nu$ for the
\asnmr central line in the $x=$ 0.068 sample. The solid lines are
fits to a Curie-Weiss law (see text). Inset: Inverse of the
temperature dependent component of the line width. The intercepts
of the linear fits with the $x$ axis correspond to $-\theta $ (see
text) . The arrows indicate $T_c$ for each radiation dose. }
\label{fwhm6}
\end{figure}
\begin{figure}[h]
\includegraphics[width=8.6cm,
keepaspectratio]{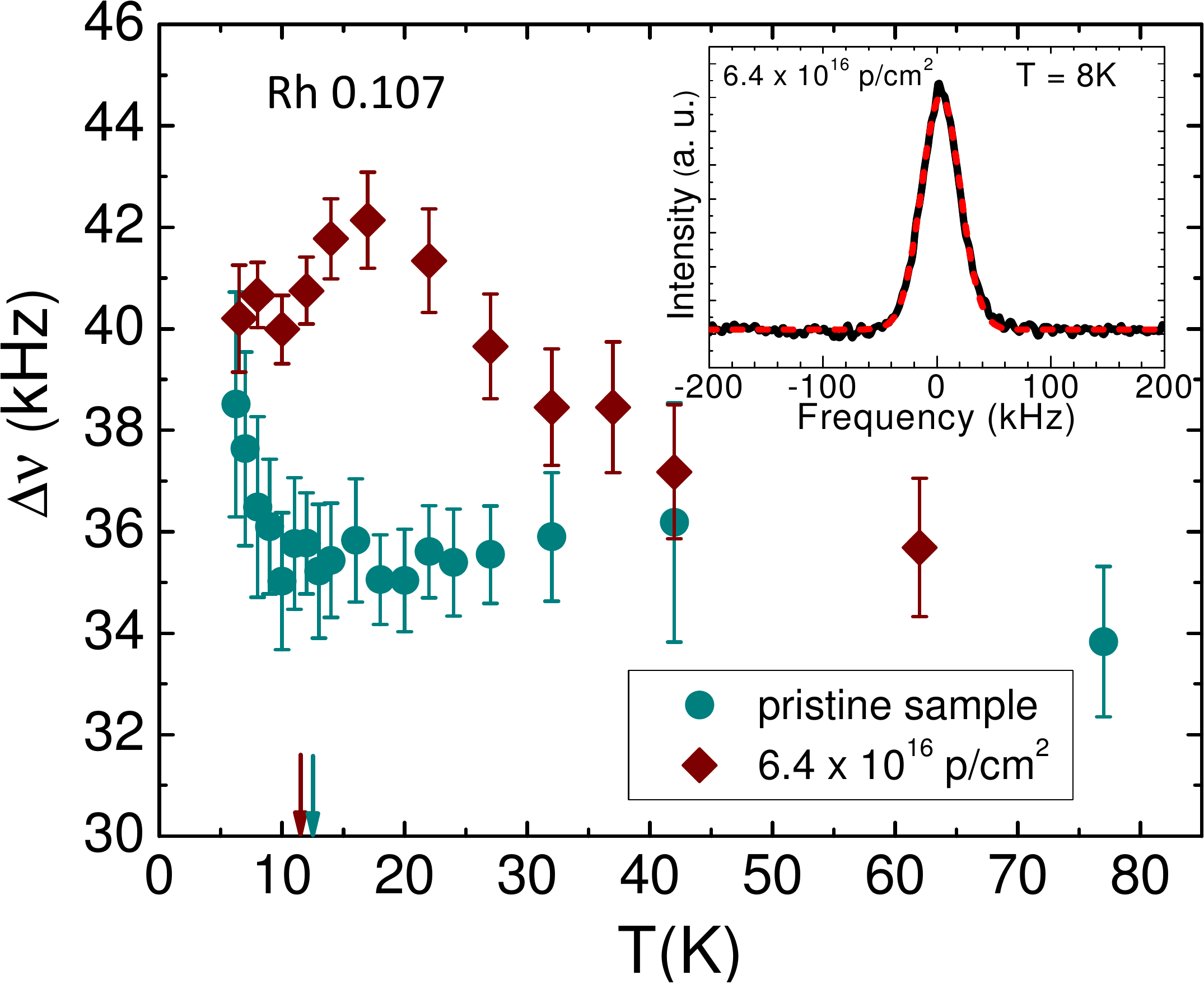} \caption{ (Color online) Temperature
dependence of the linewidth (FWHM) for the \asnmr central line in
the $x=$0.107 sample. In the inset a low temperature  \asnmr NMR
spectrum is shown, the dashed line is a fit to a gaussian
function. The arrows indicate $T_c$ for each radiation dose. The
line width data for the $\phi=$3.2$\times 10^{16}$ cm$^{-2}$ dose
level are pretty similar to those for $\phi=$6.4$\times 10^{16}$
cm$^{-2}$ and have not been reported to improve the figure
readability. } \label{fwhm10}
\end{figure}

The full width at half maximum ($\Delta\nu$ hereafter) of the
$^{75}$As central line ($m_I=\frac{1}{2}\rightarrow-\frac{1}{2}$)
was derived from the Fast Fourier Transform of half of the echo
signal after a standard Hahn spin-echo ($\pi/2-\tau-\pi$) pulse
sequence. The results for the optimally doped sample are shown in
Fig.~\ref{fwhm6} and those for the overdoped crystal can be found
in Fig.~\ref{fwhm10}. In the x=0.068 sample the linewidth
increases significantly on cooling, following a Curie-Weiss law
for all doses. Conversely, for x=0.107, $\Delta\nu$ remains nearly
flat down to T$_c$ in the non-irradiated sample while it slowly
increases, reaching a maximum around 20 K, in the irradiated one.
These strikingly different $\Delta\nu$ behaviors will be discussed
in the next section.

\begin{figure}[t]
\includegraphics[width=7.2cm]{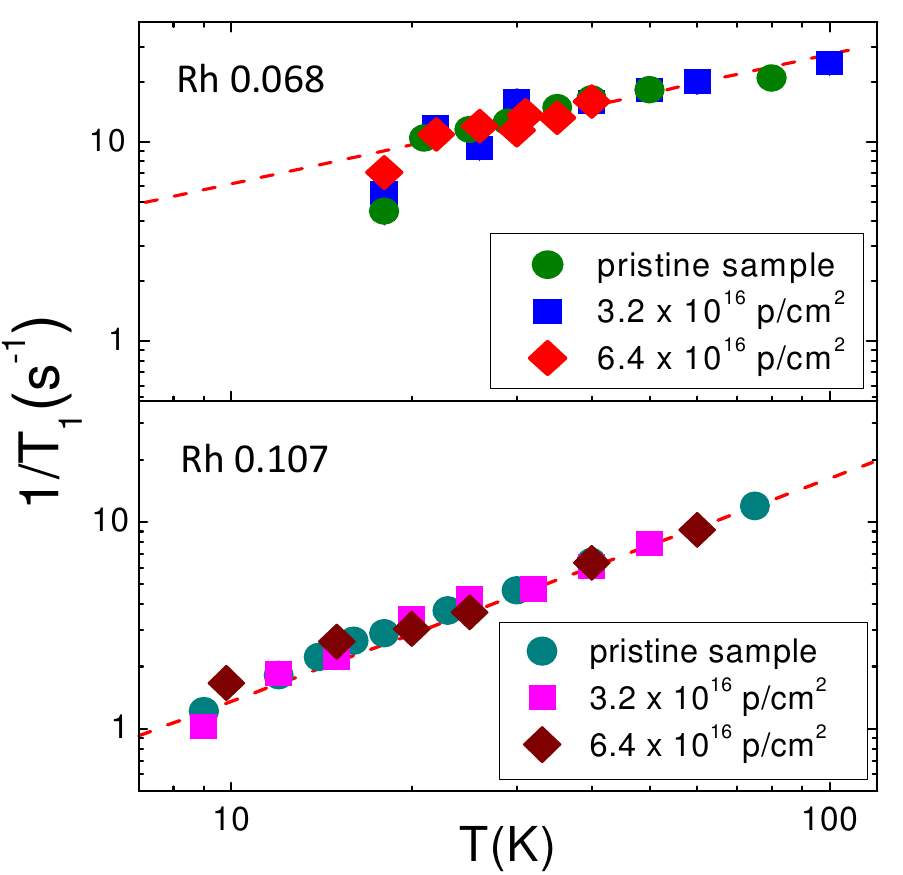} \caption{(Color online) Temperature
dependence of the \asnmr \tsl measured with H $\parallel c$  for
the $x=$ 0.068 (top) and $x=$ 0.107 (bottom) samples. The red dashed
lines are guides to the eye.}
\label{t1}
\end{figure}

The $^{75}$As spin-lattice relaxation rate was estimated by
fitting the recovery of the longitudinal magnetization $M_z(t)$
after a saturation recovery pulse sequence
($\pi/2-\tau-\pi/2-\tau_{echo}-\pi$) with the standard recovery
function for the central line of a spin 3/2 nucleus:
\begin{equation}
M_z(\tau)=M_0[1-f\, (0.1\cdot e^{-(\tau/T_1)}+ 0.9\cdot
e^{-(6\tau/T_1)})]\;\mbox{.}
\end{equation}
The results, displayed in Fig.~\ref{t1}, clearly show that 1/T$_1$
is unaffected by the presence of proton induced defects. In
particular, the spin-lattice relaxation follows a power-law
1/T$_1\sim$T$^\alpha$, with $\alpha\simeq 0.6$ for the $x=$0.068
sample and $\alpha\simeq 1$ for the $x=$ 0.107 sample, namely
close to the Korringa behavior expected for a weakly correlated
metal.

\begin{figure}[t]
\includegraphics[width=8.5cm,
keepaspectratio]{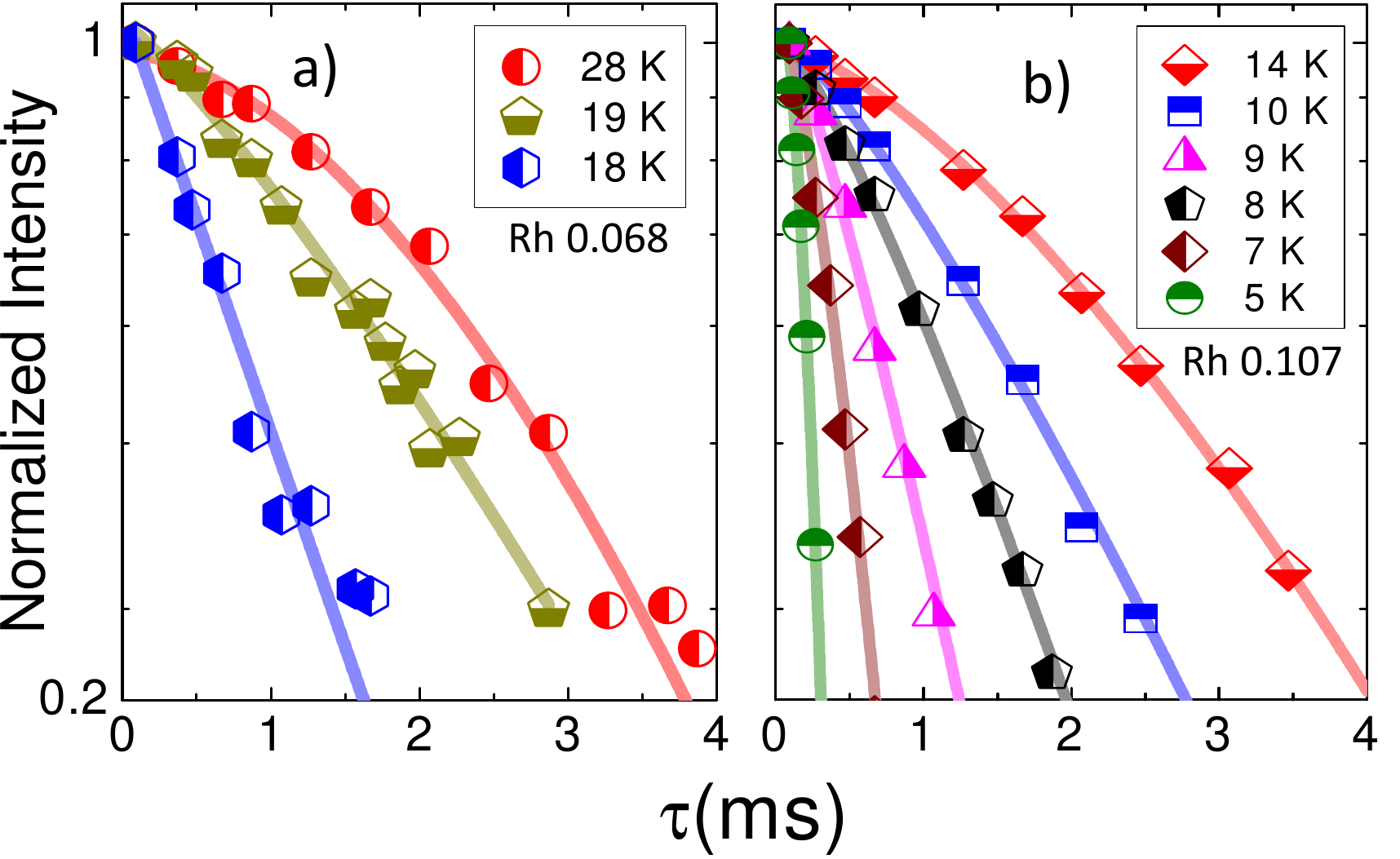} \caption{ Spin Echo decay amplitude
$M(2\tau)$ for the optimally doped $x=$ 0.068 (a) and over-doped
$x=$ 0.107 (b) samples irradiated with a fluence of 6.4$\times
10^{16}$ cm$^{-2}$, after dividing for Redfield contribution (see
text). The solid lines are fits to a stretched exponential decay
function (see text). }\label{echos}
\end{figure}

\begin{figure}[t]
\includegraphics[width=8.6cm,
keepaspectratio]{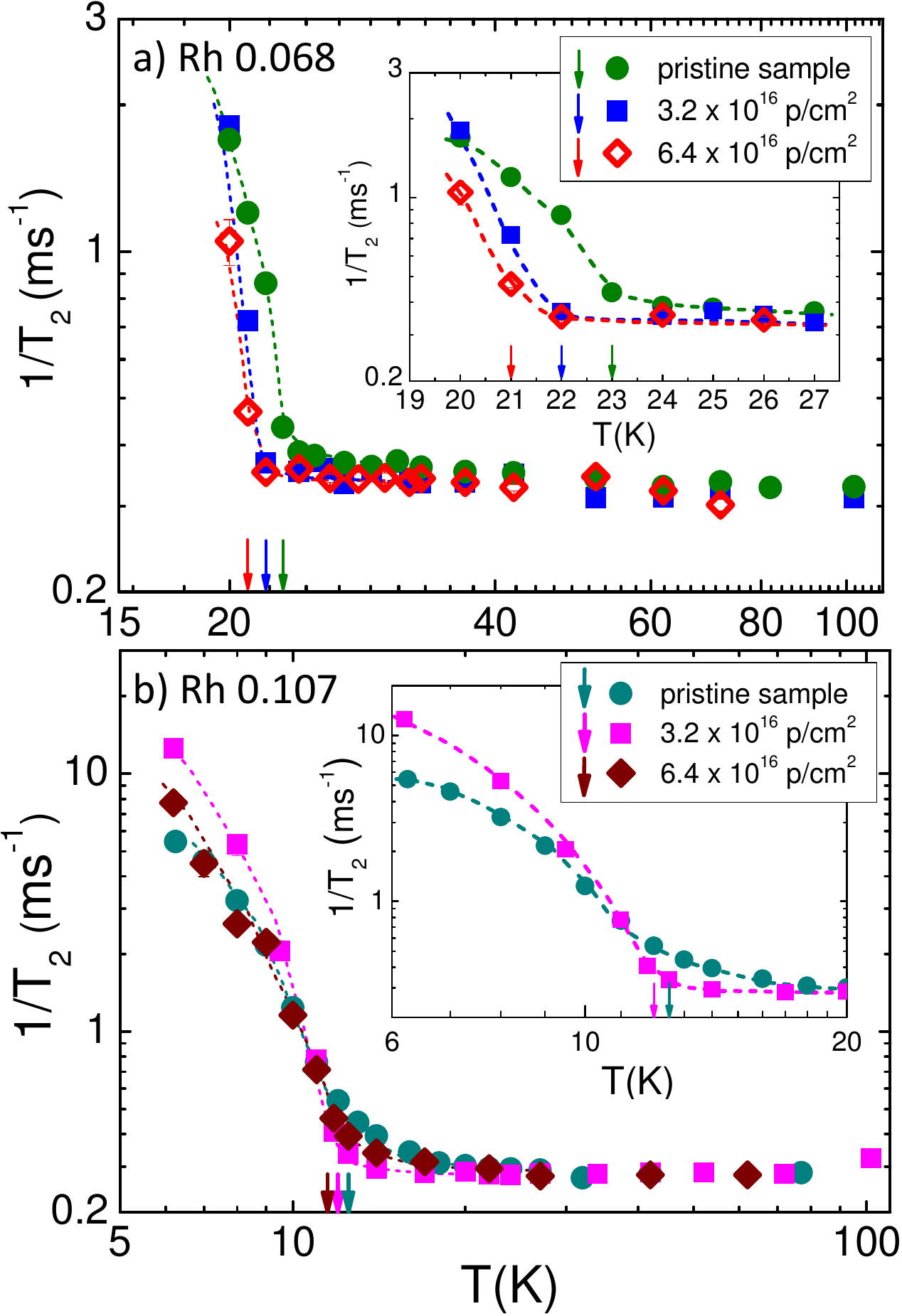} \caption{ (Color online) Temperature
dependence of the \asnmr \tss relaxation rate of the $x=$ 0.068
(top) and $x=$ 0.107 (bottom) samples for different values of
fluence (see legend).  In the insets the low temperature data are shown in
greater detail. The arrows indicate $T_c$ for each radiation dose
and Rh doping level. The dashed lines are guides to the eye.} \label{t2}
\end{figure}

The spin echo decay rate (1/T$_2$) was evaluated by recording the
decay of the spin-echo amplitude $M_{total}(2\tau)$ after a spin
echo pulse sequence. Since at high temperatures the values of T$_1$
and T$_2$ are in the same range (5-100 ms), the T$_1$ contribution
to the spin echo decay is not negligible (Redfield
term~\cite{redfield}). Within this framework the echo decay
amplitude $M_{total}(2\tau)$ can be written as~\cite{walstedt}:
\begin{equation}\label{red-echo}
M_{total}(2\tau)=M(2\tau)\exp \left(-\dfrac{2\tau}{T_{1R}}\right)
\end{equation}
where $M(2\tau)$ is the T$_1$ independent echo decay amplitude
while the exponential term takes into account the T$_1$
contribution. Walstedt and coworkers~\cite{walstedt} found that,
for the central line of a $3/2$ spin nucleus, $1/T_{1R}$ is:
\begin{equation}
\dfrac{1}{T_{1R}}=\dfrac{3}{T_1^\parallel}+\dfrac{1}{T_1^\perp}\; ,
\end{equation}
where $T_1^\parallel$ and $T_1^\perp$ denote the spin lattice
relaxation rate measured with the static magnetic field parallel
and perpendicular to the crystallographic $c$ axis, respectively.
The raw echo amplitude was then divided by $\exp
(-\frac{2\tau}{T_{1R}})$ in order to derive $M(2\tau)$. It was
found that $M(2\tau)$ deviates from a single exponential decay
(see Fig.~\ref{echos}) and could be fitted, over the whole
temperature range, by a stretched exponential:
\begin{equation}
M(2\tau)=M_0\exp\left(-\left(\dfrac{2\tau}{T_2}\right)^\beta\right)\; ,
\end{equation}
with $\beta$ the stretching exponent. The values of $\beta $ are
strongly temperature dependent (Fig.~\ref{echos}): at high
temperature $\beta\simeq 2$, indicating a Gaussian decay of the
spin echo, while it gradually decreases upon lowering the
temperature, reaching $\beta\simeq 1$ (simple exponential) close
to T$_c$. The temperature dependence of 1/T$_2$ upon varying the
dose and Rh doping is displayed in Fig.~\ref{t2}. While at
temperatures much higher than $T_c$ the spin echo decay rate is
flat for both compounds, a sharp rise in 1/T$_2$ was observed just
above T$_c$. This effect has already been reported in previous
studies (see Refs.~\onlinecite{lucia2013,lucia2016}) and is
clearly decoupled from T$_c$. In fact, by increasing the static
magnetic field~\cite{lucia2016} it is possible to shift the
1/T$_2$ increase to much higher temperatures. As it can be seen in insets of Fig.~\ref{t2}, the 1/T$_2$ upturn becomes
sharper in the proton irradiated samples. If we define $T^*$ as
the temperature below which $T_2(100 K)/T_2 > 1$, one can observe
that in both samples $T^*$ decreases upon proton irradiation.

%

\section{\label{sec:discussion} Discussion}

%
\begin{table}[b]
	\caption{Curie constant $C$ and Curie-Weiss temperature $\theta$
		obtained from the analysis of the temperature evolution of the
		$^{75}$As NMR central line width $\Delta\nu$ shown in Fig.~\ref{fwhm6} for
		Ba(Fe$_{0.932}$Rh$_{0.068}$)$_2$As$_2$. The temperature independent term
		$\Delta\nu_0$ is equal to 21.5 kHz.
	}
	\label{curie}
	\begin{ruledtabular}
		\begin{tabular}{c c c }
			$\phi$ (cm$^{-2}$) & $C$ (kHz$\cdot$K) & $\theta$ (K) \\
			\hline
			0 & $420\pm40$ & $20\pm4$  \\
			3.2$\times 10^{16}$ & $460\pm50$ & $5\pm3$  \\
			6.4$\times 10^{16}$ & $440\pm40$ & $-6.5\pm1.5$  \\
		\end{tabular}
	\end{ruledtabular}
\end{table}

Let us first consider the rich phenomenology displayed by the
$^{75}$As NMR line width (Figs.~\ref{fwhm6} and~\ref{fwhm10}). In
the optimally doped sample (x=0.068) $\Delta\nu$ increases at low
temperature for all the dose levels (see Fig.~\ref{fwhm6}).
Conversely, $\Delta\nu$ is flat at high temperature (T$>$60 K) and
its value is only weakly dependent on the total proton fluence. 
In the former compound it
is possible to fit the line width temperature dependence with a
Curie Weiss law:
\begin{equation}
\Delta\nu = \Delta\nu_0 + \dfrac{C}{T+\theta}
\end{equation}
where $\Delta\nu_0$ is a temperature independent component, $C$ is
the Curie constant and $\theta$ the Curie-Weiss temperature. The
fit parameters are summarized in Table.~\ref{curie}. The Curie
Weiss behavior of the linewidth and the observation that for T$<
50$ K $\Delta\nu$ decreases upon decreasing the magnetic field
intensity indicate that the low temperature broadening is
associated with the modulation of the local magnetic field at the
nuclei induced by the electron spin texture.

\begin{figure}[t]
\includegraphics[width=7cm,
keepaspectratio]{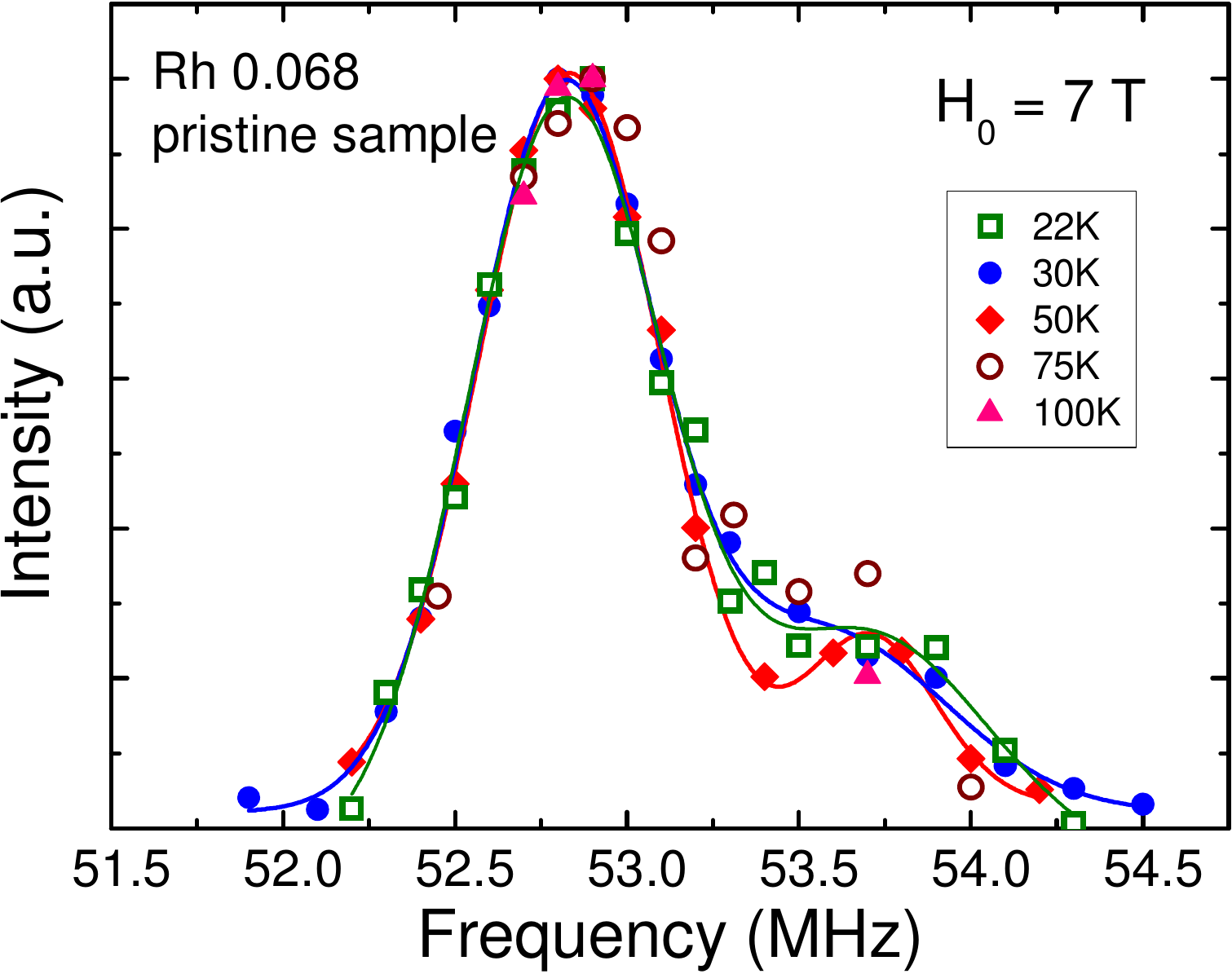} \caption{(Color online) Spectra of the high
frequency satellite line of Ba(Fe$_{0.932}$Rh$_{0.068}$)$_2$As$_2$
for various temperatures. The intensity is the integral of the
spin echo and the solid lines are fits to a double Gaussian. }
\label{satellite}
\end{figure}

The high temperature line width, $\Delta\nu_0\simeq 21.5$ kHz, is
due to the sum of nuclear dipolar line broadening, of the
quadrupolar broadening and possibly of the magnetic broadening
($\Delta\nu_{magnetic} \propto M(T,H_0) \propto \chi(T)H_0$). From
dipolar sums it can be found that the nuclear dipolar contribution
is actually very small ($<$ 2 kHz)~\cite{canfield2009,lucia2013}.
The quadrupolar broadening should be zero for $H\parallel c $,
however the misalignment by an angle $\vartheta$ may lead to some
broadening of the central $^{75}$As NMR line, which can be
estimated from \cite{abragam}
\begin{equation}
\Delta\nu_{0_Q}\sim \dfrac{3\nu_{Q}\Delta\nu_{Q}}{\nu_L} \vartheta^2 \, ,
\end{equation}
where $\nu_{Q}$ is the splitting between the central line
($\frac{1}{2}\rightarrow\frac{-1}{2}$) and the satellite line
($\frac{1}{2}\rightarrow\frac{3}{2}$), $\Delta\nu_{Q}$ the width
of the satellite, $\nu_L=\gamma H_0/2\pi$ the Larmor frequency.
The spectrum of the Ba(Fe$_{0.932}$Rh$_{0.068}$)$_2$As$_2$ high
frequency satellite line is reported in Fig.~\ref{satellite}. If
one considers that the misalignment $\vartheta< 10^{\circ}$ one
finds that the quadrupolar broadening $\Delta\nu_{0_Q}\leq$ 10
kHz, still much smaller than $\Delta\nu_0$. It is then likely that
the temperature independent magnetic broadening has to be
associated with the T-independent component of the electron spin
susceptibility, similarly to what reported by Mukhopadhyay et
al.~\cite{canfield2009} in Ba$_{1-x}$ K$_x$Fe$_2$As$_2$.

The Curie-Weiss $\Delta\nu$ behavior indicates the presence of
spin correlations and  was often observed in the cuprates in the
presence of defects\cite{alloul2009}. In fact, the impurities
induce a local spin polarization $\langle S_z \rangle$ on the
conduction electrons  which leads to a spatially varying spin
polarization $s(\mathbf{r})=\chi (\mathbf{r})\langle S_z \rangle$.
The resulting NMR spectrum is the histogram of the spin
polarization probed by the nuclei and the line width at a given
temperature depends on the temperature evolution of $\chi
(\mathbf{r})$. Accordingly, $\Delta\nu$ follows the susceptibility
of the local moments which can be described by a Curie-Weiss
law~\cite{alloul2009}.

Remarkably, for \mbox{$\phi=6.4\times 10^{16}$~cm$^{-2}$}, the
Curie-Weiss temperature becomes negative, signaling the shift of
the correlations from antiferromagnetic to ferromagnetic.
Ferromagnetic correlations were detected in other compounds of the
122 family, in particular in the non-superconducting
Ba(Fe$_{1-x}$Mn$_x$)$_2$As$_2$~\cite{leboeuf2014} and, with a much
lower $\theta$, in the superconducting
Ba(Fe$_{1-x}$Co$_x$)$_2$As$_2$ and
BaFe$_2$(As$_{1-x}$P$_x$)$_2$~\cite{texier2012} after the
introduction of Mn impurities. Ferromagnetic fluctuations were also observed in 
hole and electron doped BaFe$_2$As$_2$ \cite{FurukawaPRL} and in 
Ca(Fe$_{1-x}$Co$_x$)$_2$As$_2$ \cite{FurukawaCa122}. The observed $\Delta\nu$
temperature dependence is analogous to the one measured in
Mn-doped LaFe$_{1-y}$Mn$_y$AsO$_{1-x}$F$_x$, where the
introduction of tiny amounts of Mn strongly suppresses T$_c$ and
gives rise to a significant increase of $^{19}$F NMR line
width\cite{hammerath2015,moroni2016}. However, in
LaFe$_{1-y}$Mn$_y$AsO$_{1-x}$F$_x$, $\theta$ is always positive
and the introduction of magnetic impurities enhances both $\theta$
and $C$, indicating  that Mn doping strengthens the spin
correlations already present in the Mn free compound. On the other
hand, in proton irradiated Ba(Fe$_{0.932}$Rh$_{0.068}$)$_2$As$_2$,
the value of $C$ remains unchanged and $\theta$ first decreases
and then changes sign upon increasing the dose.

It should be noticed that, at variance with Mn doping, the lattice
defects created by proton irradiation are nonmagnetic.
Even though the rise of magnetism upon ion irradiation was
observed in several materials ~\cite{esquinazi2013} and we recall
that the Ba122 family of iron-based superconductors is quite
unstable towards impurity driven static magnetism
~\cite{canfield2009,goko2009}. Hence, the observation that the
nonmagnetic defects introduced by irradiation lead to enhanced
spin correlations and to a broadening of the NMR lines is not
unexpected. Indeed, it is well known that by doping
YBaCu$_{3}$O$_{6+x}$ with nonmagnetic Zn impurities the $^{89}$Y
NMR line gets structured~\cite{alloul91}
 and its line width follows a Curie
law~\cite{mahajan94}.

In the overdoped sample the behavior of the linewidth is
completely different from that of the optimally doped (see
Fig.~\ref{fwhm10}). The pristine sample displays a completely flat
$\Delta\nu (T)$ down to 9 K and then a rapid increase, likely due
to the freezing of the vortex motions~\cite{oh2011}. In the
irradiated sample (\mbox{$\phi=6.4\times 10^{16}$~cm$^{-2}$})
$\Delta\nu$ reaches a maximum around 18 K and then decreases
slightly at lower temperatures. Interestingly the temperature at
which the line width of the irradiated sample starts to decrease
is very near to the temperature T$^*$ at which the spin-spin relaxation
rate starts to rise and the echo decay becomes a single
exponential. This suggests that the low frequency spin
fluctuations, which are responsible for the \tss enhancement,
partially average out the static frequency distribution probed by
the $^{75}$As nuclei.

We will now discuss the effect of irradiation on \tss . The marked
increase of \tss starting at T$^*>$ T$_c$ seems to be a common
feature of several 122
compounds~\cite{oh2011,lucia2016,lucia2013}. In
Ref.~\onlinecite{lucia2016} we showed that this effect is
unrelated to the superconducting state and that T$^*$ can become
much higher than T$_c$ in the presence of a high magnetic field. As we
already explained in the previous section the \tss enhancement
below T$^*$ is affected by proton irradiation. The increase in
\tss was associated with slow nematic fluctuations between ($\pi$,0) and (0,$\pi$) ground states, very much akin to the nematic fluctuations found in
prototypes of the J$_1$-J$_2$ model on a square
lattice.~\cite{carretta2002,chandra1990} These low-frequency fluctuations have
been predicted~\cite{mazin2009} in the iron based superconductors
and nematic fluctuations have subsequently been observed in
several underdoped~\cite{fu2012,iye2015,zhou2016} and overdoped
IBS~\cite{lucia2013,lucia2016}.

In the presence of these fluctuations \tss can be written as:
\begin{equation}\label{model-t2}
\dfrac{1}{T_2}= a (\Delta\nu(T))^2\tau_D(T)+ \dfrac{1}{T_{2_i}}
\end{equation}
with $\tau_D$ the characteristic fluctuation time, $a$ a
dimensionless coupling constant and T$_{2_i}$ the T-independent
contribution to the relaxation arising from nuclear dipole-dipole
interaction. The resulting temperature dependent $\tau_D(T)$ can
then be fitted to an Arrhenius law $\tau_D(T)=\tau_0e^{U/T}$ where
$U$ is the activation energy and $\tau_0$, the high temperature
characteristic time of the fluctuations, in the nanosecond range. We fitted the \tss data using Eq.~\ref{model-t2} in the 20 - 26 K temperature range for x=0.068 and in the 7 K - 30 K range for x=0.107.

In the pristine samples we found that, for x=0.068, the activation
energy is $U\simeq200\pm 30$ K while in the overdoped $x=$0.107
sample $U\simeq40\pm 20$ K, in good agreement with the values
obtained in Ref.~\onlinecite{lucia2013,lucia2016}. Upon proton
irradiation $U$ increases markedly in the optimally doped sample
($U\sim 500 \pm 100$ K for \mbox{$\phi=3.2\times
10^{16}$~cm$^{-2}$}) while it remains basically unchanged in the
overdoped sample. Unfortunately, the quality of the fit decreases
with increasing dose, pointing out that possibly the dynamics can
no longer be described by a single activation barrier and that a
distribution of energy barriers should be considered. This fact is
particularly evident in the overdoped sample where the increase of
\tss becomes significantly sharper and T$^*$ decrease by $\sim$ 5
K (Fig. 6). The substantial enhancement of the activation energy
suggests that the presence of the defects slows down the fluctuations between the
$(0,\pi)$ and ($\pi$,0) ground states. It is remarked that
such an effect has also been detected in the prototypes of the
J$_1$-J$_2$ model on a square lattice doped with nonmagnetic
impurities.\cite{papinutto2005}

\section{\label{sec:conclusion} Conclusions}
In conclusion, we have shown that proton irradiation (5.5 MeV) in
\bafeas results in a very weak T$_c$ suppression, in good
agreement with previous experiments carried out in other 122
compounds~\cite{nakajima2010, taen2013}. By measuring the
$^{75}$As NMR spectra we have evidenced that the defects
introduced by proton irradiation induce ferromagnetic correlations
in the optimally electron doped x=0.068 compound. Remarkably this
effect is totally absent in the overdoped sample owing to the
absence of significant spin correlations. Moreover the analysis of
the spin echo decay rate (1/T$_2$) show that the low-frequency
fluctuations
observed~\cite{xiao2012,curro2009,fu2012,iye2015,zhou2016} in
several families of iron based superconductors are damped by the
irradiation induced impurities, consistently with the hypothesis
that they could be associated with the presence of nematic
fluctuations between ($0,\pi)$ and ($\pi$,0) nematic phases.

\section*{acknowledgments}
This work was supported by MIUR-PRIN2012 Project No. 2012X3YFZ2
and MIUR-PRIN2015 Project No. 2015C5SEJJ. The irradiations were
performed in the framework of the INFN-Politecnico di Torino
M.E.S.H. Research Agreement. The work at the Ames Laboratory was supported by 
the DOE-Basic Energy Sciences under Contract No. DE-AC02-07CH11358.The authors wish to thank the
INFN-LNL staff for their support in the irradiation experiments.

\end{document}